\documentclass[%
 aip,
 amsmath,amssymb,
 reprint,%
]{revtex4-1}

\usepackage{graphicx}
\usepackage{dcolumn}
\usepackage{bm}

\usepackage[utf8]{inputenc}
\usepackage[T1]{fontenc}
\usepackage{mathptmx}
\usepackage{etoolbox}
\usepackage{bigints}
\usepackage[dvipsnames]{xcolor}
	\definecolor{newblue}{RGB}{0, 94, 184}
\usepackage{comment} 
\usepackage{multirow}
\usepackage[export]{adjustbox}
\usepackage[section]{placeins}
\makeatletter
\def\@email#1#2{%
 \endgroup
 \patchcmd{\titleblock@produce}
  {\frontmatter@RRAPformat}
  {\frontmatter@RRAPformat{\produce@RRAP{*#1\href{mailto:#2}{#2}}}\frontmatter@RRAPformat}
  {}{}
}%
\makeatother

\begin{document}

\preprint{}

\title{Bulk versus interface nucleation of CO$_2$ hydrates from computer simulations}

\author{Joanna Grabowska*}
\affiliation{Department of Physical Chemistry, Faculty of Chemistry, Gdansk University of Technology, ul. Narutowicza 11/12, 80-233 Gdansk, Poland}

\author{Samuel Blazquez}
\affiliation{Dpto. Qu\'{\i}mica F\'{\i}sica I, Fac. Ciencias Qu\'{\i}micas, Universidad Complutense de Madrid, 28040 Madrid, Spain}

\author{Carlos Vega}
\affiliation{Dpto. Qu\'{\i}mica F\'{\i}sica I, Fac. Ciencias Qu\'{\i}micas, Universidad Complutense de Madrid, 28040 Madrid, Spain}

\author{Eduardo Sanz$^*$}
\affiliation{Dpto. Qu\'{\i}mica F\'{\i}sica I, Fac. Ciencias Qu\'{\i}micas, Universidad Complutense de Madrid, 28040 Madrid, Spain}

\begin{abstract}

\textbf{Abstract}

Gas hydrates are of great relevance to both the oil industry and the environment. Understanding how these solid structures nucleate from aqueous solutions is essential to controlling their formation. Experimental studies have often suggested that hydrate nucleation originates at the interface between the aqueous phase and the guest-molecule reservoir. To assess this hypothesis, we perform molecular dynamics simulations of CO$_2$ hydrate nucleation. First, we place hydrate seeds at different positions relative to the interface and monitor their evolution, finding that seeds embedded in the bulk are more likely to grow   than those located near or at the interface. Second, we analyse spontaneous nucleation simulations with and without an interface. Our previous work showed that nucleation rates are indistinguishable in both systems, strongly indicating that the interface does not play a role. Here, trajectory analysis reveals that hydrates nucleate in regions of locally high CO$_2$ concentration, which arise spontaneously in the bulk and are not associated with the interface. Our results indicate that hydrate nucleation does not preferentially occur at the interface, at least at the
at deep supercooling conditions explored in this work. 
Further work at higher temperatures, and considering alternative nucleation locations, 
is needed to reconcile experiments and simulations, and thereby reach a deep
understanding of the mechanism of hydrate formation. 

\end{abstract}

\maketitle
$^*$Corresponding authors: \\
J. Grabowska: joanna.grabowska@pg.edu.pl,\\
E. Sanz: esa01@quim.ucm.es

\section{Introduction}

Clathrate hydrates are crystalline structures in which gas molecules are trapped within cages of water molecules.\cite{Sloan2008a} 
Methane hydrates are primarily found in deep-sea sediments and permafrost regions, and serve as a substantial reservoir of methane and other fuel molecules.
In addition to their potential as an energy source, gas hydrates offer a promising avenue for carbon dioxide (CO$_2$) sequestration, as CO$_2$ could replace methane within hydrate structures, \cite{hirohama1996conversion,park2006sequestering,lee2003recovering,lee2013experimental} providing a dual benefit of reducing greenhouse gas emissions while facilitating methane extraction. 
Hydrogen hydrates are also important because they offer a potential method for safe and compact storage of hydrogen, supporting the transition to cleaner energy sources.
Methane hydrates, however, pose significant challenges for the oil and gas industry, particularly due to their tendency to form blockages in pipelines during hydrocarbon production and transportation, leading to operational inefficiencies and safety risks.\cite{sloan2021hydrocarbon} It is thereby crucial to characterise and understand
the formation and stability of hydrates. 

Numerous experimental studies \cite{Ruoff1994, Myerson1999, Uchida2000, Cournil2004, Svartaas2010, Solms2018, Maeda2018, Maeda2019,liang2019nucleation} have been conducted to better understand the formation, growth and dissociation of gas hydrates under various thermodynamic and kinetic conditions. Researchers have investigated the effects of temperature, pressure, gas composition and the presence of inhibitors or promoters on hydrate crystallization and stability.

There seems to be compelling experimental evidence that 
 hydrates nucleate at the interface between the solution and 
 the hydrate former reservoir. \cite{maeda2015nucleation,stoporev2018visual,Sloan2008a,adamova2018visual,li2024dependence,jeong2022gas,maeda2015nucleation,sloan2007clathrate} This is a 
 likely scenario given that the solubility of
 hydrate former molecules (CO$_2$, CH$_4$, H$_2$S etc.) in water is typically much lower than the proportion of gas former molecules in the solid hydrate, which is 1 molecule per 5.75 water molecules in a perfect sI hydrate lattice. Thus, it appears reasonable that only in the vicinity of the interface there is enough hydrate former concentration to nucleate the hydrate. 
Experimental techniques, however, do not enable a direct visualization of the
first hydrate
embryo that nucleates from a disordered molecular arrangement, given its nanoscopic size and its short lifespan.

A theoretical approach is very useful for understanding and rationalizing
the competition between homogeneous and interfacial
nucleation; however, the values of the relevant parameters
that enter the theoretical description require experimental
validation, and the theory does not provide a molecular-level
representation of the nucleation process.\cite{Kashchiev2002b}
Molecular simulations can help bridge the gap in our
understanding of nucleation at a molecular scale,
\cite{Clancy1994, Smith1996, Alavi2005a,Alavi2006a,Wu2009, English2009a,Walsh2011a, Sarupria2012b, Knott2012a, Liang2013a, Barnes2014a, Yuhara2015a, Zhang2016b, Lauricella2017a, Arjun2019a, Arjun2020a, Arjun2021a, Arjun2023,Liang2013a,Zhang2018a,Zhang2020a,Wang2003a,Jimenez-Angeles2014a,Jimenez-Angeles2018a,Zhang2022a,Tanaka2023b,Tanaka2024a} particularly so with 
the emergence of 
realistic water models that have proven to be highly accurate in predicting the
equilibrium behaviour of real hydrates.\cite{Conde2010a,Miguez2015a,blazquez2024three,algaba2024three,Waage2017a}
Most simulation studies of hydrate nucleation, however, focus on the homogeneous nucleation 
of the hydrate in the bulk aqueous solution \cite{Knott2012a,Yuhara2015a,Grabowska2022b,zeron2025homogeneous,Arjun2020a,Arjun2021a,li2020unraveling,Wu2009,Sarupria2012b} (in many cases with a guest molecule concentration
way higher than 
the saturation concentration to enhance the nucleation rate) and the 
hypothetical location of the nucleus at the interface 
has not been properly assessed. 

Some clues regarding the location of hydrate nucleation can be found in Refs. 
\onlinecite{Jacobson2010a} and \onlinecite{zhang2025temperature}.
In Ref. \onlinecite{Jacobson2010a}, focused on unveiling the molecular path leading to the formation of hydrates
(blobs of guest molecules give rise to hydrate nuclei),
it was mentioned that nuclei can indistinctly appear 
in the bulk 
or at the interface.
According to Ref. \onlinecite{zhang2025temperature},
in contrast, nucleation in the bulk is not considered as a possibility (it either takes place at the interface with the 
hydrate-former rich phase or with a solid substrate
at low and high temperatures respectively). 

The fact that homogeneous nucleation cannot compete with nucleation at the interface 
is inconsistent with our recent simulation study. \cite{zeron2025homogeneous} 
In Ref. \onlinecite{zeron2025homogeneous} we conducted simulations
of spontaneous 
CO$_2$ hydrate nucleation from a bulk CO$_2$ saturated solution and from a 
CO$_2$ saturated solution in contact with a CO$_2$
reservoir through a flat interface.
The frequency of hydrate nucleation was the same in both cases, suggesting that the interface does not promote nucleation, at least under the studied conditions (245 and 250 K, 400 bar).

In this work, we use Molecular Dynamics (MD) simulations to clarify whether hydrates preferentially appear at the solution-hydrate former interface. More specifically, 
we place CO$_2$ hydrate seeds at different locations relative to the CO$_2$-solution interface
and track their evolution in constant pressure, constant temperature ($NpT$) MD simulations.
We find that, in the conditions of our study (400 bar and 255 K), proximity to the interface hinders the growth of CO$_2$ hydrate seeds, leading to faster nucleation rates in the bulk than at the interface.

Our work highlights that hydrate nucleation in the bulk aqueous solution is faster
than at the interface with the hydrate former. 
This result seems to be at odds with the general understanding that nucleation is faster at the interface. \cite{maeda2015nucleation,stoporev2018visual,Sloan2008a,adamova2018visual,li2024dependence,jeong2022gas,maeda2015nucleation,sloan2007clathrate}
A study at higher temperatures (closer to the dissociation temperature, where experiments
are typically carried out) is needed to explore the possibility of a crossover of nucleation 
location along temperature. Also, nucleation at the container walls or
assisted by impurities are possibilities that could reconcile our results with the
widely visualised observation of the appearance of hydrates at the interface in 
experimental research.

\section{Methodology}

\subsection{Simulation Details}

Water and CO${_2}$ molecules are modeled with TIP4P/Ice\cite{Abascal2005b} and TraPPE,\cite{Potoff2001a} respectively. Water–CO${_2}$ dispersive interactions are treated with the modified Lorentz-Berthelot rule proposed by Míguez \emph{et al.} (which simply consists in multiplying by 1.13 the cross energy parameter given by Lorentz-Berthelot).\cite{Miguez2015a} These potentials yield accurate predictions of the CO${_2}$ hydrate three-phase line, with the dissociation temperature $T_{3}$ at 400 bar matching the experimental value (286 K in experiment versus 290 K in simulations \cite{Algaba2023a}). The model, however, 
 is not perfect. For instance, it overestimates the solubility of CO$_2$
 in liquid water at low temperatures \cite{Blazquez2024four}. Importantly, while this force-field limitation affects the absolute value of the CO$_2$ solubility, it does not alter the qualitative conclusions of the present work. All simulations were performed using the same force field, and comparisons between bulk and interfacial nucleation were therefore made under internally consistent conditions.

We focus on a pressure of 400 bar and 
most simulations are carried out at 255 K
(i.e. 35 K supercooling). 
We focused on a pressure value typical of experiments in order to make predictions at experimentally relevant conditions and to
enhance the solubility of CO$_2$ in water. Other pressure values should be
investigated to properly assess the role of 
this variable, that may affect the nucleation rate and pathway.\cite{Kashchiev2002b}

MD simulations were used to obtain all the results presented in this work. The simulations were conducted with the use of the GROMACS package. \cite{VanDerSpoel2005a,hess2008gromacs} The leapfrog algorithm with a time step of 2 fs has been used to integrate the equations of motion. Temperature was kept constant with the use of a Nos\'e-Hoover \cite{Nose1984a,Hoover1985a} thermostat with a relaxation time of 2 ps, while pressure was kept constant with the Parrinello-Rahman \cite{Parrinello1981a} barostat with the same relaxation time. The isothermal-isobaric ensemble (Np$_x$T) has been used for most of the runs (with the barostat applied only along the $x$ direction, which is normal to the interface); in some cases, NVT simulations were employed (details of the specific simulations will be provided below). For dispersive and Coulombic interactions, a cut-off of 1 nm has been used. Particle Mesh Ewald method \cite{Darden1993} has been used to compute electrostatic interactions. Long-range corrections for dispersive interactions were not used in our simulations.

All simulation systems were prepared using a two-phase system, where an aqueous solution of CO$_2$ molecules was in contact - via a planar interface - with a CO$_2$ reservoir. The size of this system was 11.6 x 8.5 x 8.5 nm$^3$, where the $x$ direction is perpendicular to the interface, and it contained 15735 molecules of water and 5 896 molecules of CO$_2$. 

The initial configuration of the two-phase system was prepared by combining pre-equilibrated simulation boxes of pure CO$_2$ and 
of a CO$_2$ aqueous solution  (with CO$_2$ molar fraction of 0.085, which is close to the expected equilibrium solubility of CO$_2$ in water for the selected model under the studied conditions \cite{Algaba2023a}). The obtained system was then equilibrated for 40 ns at 255 K and 400 bar, in an anisotropic Np$_x$T ensemble (i.e., the pressure was allowed to fluctuate only in the direction perpendicular to the interface). During the equilibration run, we monitored changes of the molar fraction of CO$_2$ in the aqueous phase, which reached a stable value of approximately 0.077 after 20 ns. The equilibration run was continued for another 20 ns to ensure that the CO$_2$ concentration in the aqueous phase remained at equilibrium. 

After equilibration of the aqueous solution in contact with the CO$_2$ reservoir, the Seeding method was used: CO$_2$ hydrate crystal seeds of varying sizes were extracted from a bulk sI hydrate crystal in which all cages were fully occupied by CO$_2$ molecules (the CO$_2$ hydrate was equilibrated beforehand at 255 K and 400 bar for 10 ns) and inserted into the prepared two-phase system. After this step, 
the interface between the
inserted cluster and the surrounding fluid  
was equilibrated with the use of three different protocols as explained in Supporting Information.

\begin{figure}
\centering
\hspace*{-0.2cm}
\includegraphics[width=0.9\columnwidth]{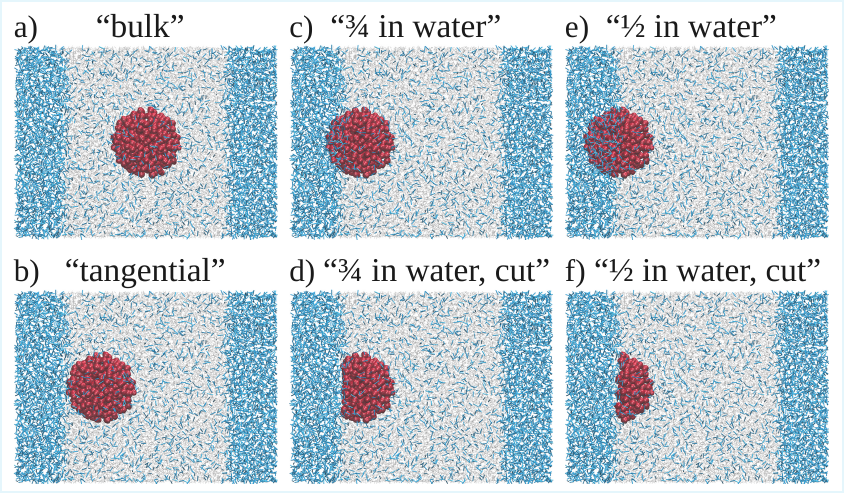}
\caption{Snapshots of the seed locations and shapes investigated in this work: a) bulk, b) tangential, c) 3/4 in water, d) 3/4 in water, cut, e) 1/2 in water, f) 1/2 in water, cut. The colour scheme of the figure is as follows: all molecules in the seed are represented as red spheres, water molecules and CO$_2$ molecules which do not belong to the seed are represented as grey and cyan sticks, respectively. The seeds presented in the figure have a radius equal to 1.5 nm. The graphic was created using VMD\cite{vmd} software.}
\label{fig:system_types}
\end{figure}

For all equilibration protocols, 6 types of systems were prepared, differing in the positioning of the seed with respect to the 
CO$_2$-aqueous solution interface. The different seed locations
are shown in the snapshots in Fig. \ref{fig:system_types}. Below, we indicate
how we refer to each of these locations and briefly describe
them:

\begin{enumerate}
    \item \emph{Bulk} (Fig. \ref{fig:system_types} (a)): the seed is inserted in the middle of the aqueous phase, which is the type of setup we used in our previous Seeding work \cite{zeron2025homogeneous}. 
  \item \emph{Tangential} (Fig. \ref{fig:system_types} (b)): the seed is inserted tangentially touching the CO$_2$-aqueous solution interface. 
  \item \emph{3/4 in water} (Fig. \ref{fig:system_types} (c)): 1/4$^{th}$ of the 
  seed diameter is immersed in the CO$_2$ liquid and the 
  remaining 3/4$^{th}$ is immersed in the aqueous solution. 
  \item \emph{3/4 in water, cut} (Fig. \ref{fig:system_types} (d)): same as
  \emph{3/4 in water} but the part of the seed 
  immersed in the CO$_2$ phase is cut off. 
  \item \emph{1/2 in water} (Fig. \ref{fig:system_types} (e)): half of the 
  seed is immersed in the CO$_2$ phase and the other half in
  the aqueous solution. 
   \item \emph{1/2 in water, cut} (Fig. \ref{fig:system_types} (f)): same as
  \emph{1/2 in water} but the part of the seed 
  immersed in the CO$_2$ phase is cut off. 
\end{enumerate}

After the equilibration period we either fix the positions of atoms in the seed and let the seed grow or calculate the probability with which the unrestrained seed grows or shrinks. We determined the starting seed size as an average within the first 3 ns of a given production run. 

\begin{figure}
\includegraphics[width=0.45\textwidth]{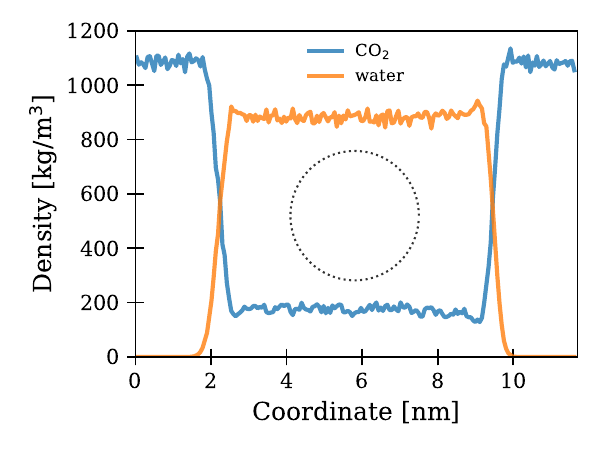}
\caption{Density profiles of CO$_2$ and water along the direction perpendicular to the CO$_2$-water interface in the system used for seeding nucleation simulations. The circle marks the approximate spatial extent of the inserted hydrate seed of a radius of 1.7 nm and is included to provide a reference length scale relative to the profile gradients.}
\label{fig:densprof}
\end{figure}

To make sure that the \emph{bulk} placement really corresponds
to a molecular environment typical of a bulk aqueous solution
we compute density profiles of CO$_2$ and water across the
interface.
Such density profiles, shown in Fig. \ref{fig:densprof}, enable a microscopic identification of the interface based on the spatial variation of water and CO$_2$ densities. As shown in the figure, deviations of the CO$_2$ concentration in the aqueous phase from its bulk value are confined to a region extending approximately 0.7-0.8 nm from the CO$_2$-rich phase. Beyond this distance, both the water density and the CO$_2$ concentration reach plateau values characteristic of bulk aqueous solution. Based on this analysis, we defined the interfacial region as the zone in which the density and composition vary continuously between the CO$_2$-rich and aqueous phases, and the bulk aqueous region as the region where these properties are spatially uniform. The hydrate seeds placed in the \emph{bulk} configuration were positioned at distances greater than the extent of the influence of the interface on CO$_2$ concentration (see the dashed circle in Fig. \ref{fig:densprof}). At these distances, the local environment is indistinguishable from bulk water in terms of density and CO$_2$ concentration.
Therefore the hydrate seeds in systems labeled as
\emph{bulk} were not influenced by interfacial perturbations, while seeds placed closer to the CO$_2$-rich phase clearly reside within the interfacial region.

It is important to note that there is a substantial disparity between the CO$_2$ concentration in the aqueous phase under the conditions studied (molar fraction $\approx$ 0.077) and the CO$_2$ content in the hydrate phase (molar fraction $\approx$ 0.15). Under such conditions, the growth of hydrate-like structures could, in principle, lead to local CO$_2$ depletion, potentially decelerating further growth unless rapid replenishment from a nearby CO$_2$-rich phase occurs. In the present work, however, we focus on the nucleation regime and on the very early stages of critical nucleus growth. Within the time scales and cluster sizes probed by our seeding simulations, the CO$_2$ concentration in the aqueous phase remains effectively constant. The inserted seeds do not grow large enough to induce measurable CO$_2$ depletion in the surrounding solution, and consequently, mass transport limitations do not significantly influence the nucleation kinetics reported here.

\section{Results}

\subsection{Simulations with hydrate seeds in fixed positions}

\begin{figure*}
\includegraphics[width=0.95\textwidth]{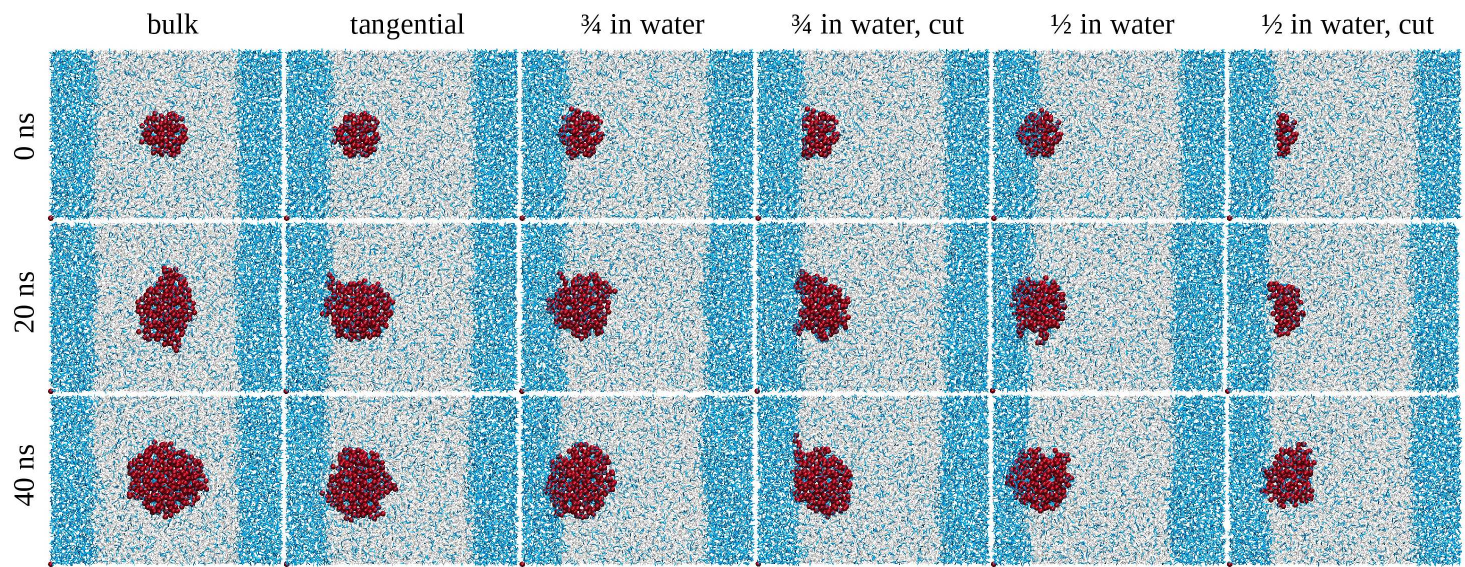}
\caption{Configurations of the systems during the seeding simulations in which all molecules in the inserted hydrate seed were kept in fixed positions. Columns correspond to different types of seed locations whereas rows correspond to different times as indicated in the figure. Water molecules that were labelled as hydrate using a linear combination of $\overline{q}_3$ and $\overline{q}_{12}$ order parameter are presented as red spheres. Molecules of water in the liquid phase and molecules of CO$_2$ are shown as gray and cyan lines, respectively. It can be observed that the seeds retain a roughly spherical shape during the simulations and those which were originally only a part of a sphere recover the spherical shape as the simulations progress. The growth of the seeds occurs mainly towards the liquid phase; in some cases, a shift of the position of the interface can be observed in order to accommodate the growth. Keep in mind that during these simulations no melting of the seeds is possible, since the molecules of the original seed inserted into the systems are kept in fixed positions during the simulations. The graphic was created using VMD \cite{vmd} software.}
\label{fig:EQ3_picture}
\end{figure*}

As described in the Methods section, we used six types of systems in our study, which differed in the position and/or the shape of the inserted hydrate seed.  To investigate the growth behaviour of the
inserted seeds we fixed the positions of all their constituent atoms during the simulations (performed at 255 K and 400 bar).
As a result, seed growth was observed in all runs, regardless of their placement. Figure \ref{fig:EQ3_picture} shows the configurations from one of the runs for each type of seed 
placement, at 0, 20 and 40 ns. The main conclusion is that seeds with an initially spherical shape maintained an approximately spherical form, whereas seeds prepared as truncated spheres at the beginning of the run tended to acquire a spherical shape during the simulation. The tendency of the seeds to acquire a spherical shape 
suggests that the scenario of cap-shaped seeds
nucleating on top of the CO$_2$-aqueous solution interface \cite{Kashchiev2002b} does not accurately represent hydrate nucleation. 

\begin{figure}
\includegraphics[width=0.9\columnwidth]{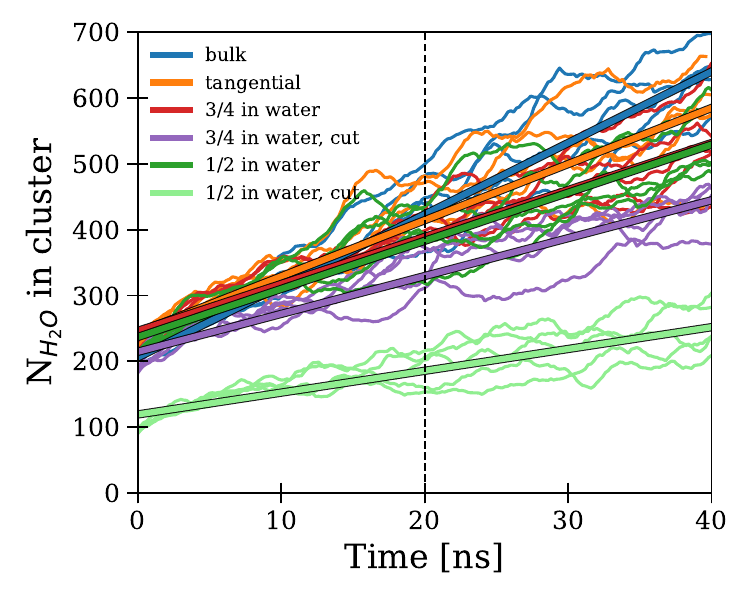}
\caption{Changes of size of the hydrate seed during the seeding simulations in which all molecules in the inserted seed were kept in fixed positions. The radius of the seeds cut from the bulk hydrate was equal to 1.3 nm. Regression lines are included to show differences in the rates of growth of the seeds. The values of the slopes are (in 1/ns unit): 10.86 for \emph{bulk}, 8.53 for \emph{tangential}, 7.14 for \emph{3/4 in water}, 5.75 for \emph{3/4 in water, cut}, 7.30 for \emph{1/2 in water} and 3.32 for \emph{1/2 in water, cut}, respectively. The configurations obtained after 20 ns of simulation (indicated in the figure as a black dashed line) were used for the production runs (EQ3 equilibration protocol, described in Supporting Information).}
\label{fig:EQ3}
\end{figure}

In Fig. 
\ref{fig:EQ3} the number of water molecules in the hydrate seed is plotted
versus time for the different seed placements under study. In order to 
quantify the seed size we count the number of molecules of water
it contains using 
a linear combination of $\overline{q}_3$ and $\overline{q}_{12}$ order parameters, as we also did in our previous work.\cite{zeron2025homogeneous}

The starting radius of all seeds was equal to 1.3 nm. 
Consequently, the starting size of the seed, measured as the number of 
water molecules it contains, is smaller in the systems where a portion of 
the sphere was cut off. 
The growth rate can be quantified through the slope of a linear fit 
to the $N_{H_2O}(t)$ curves. Such linear fits are shown with thick lines
in Fig. \ref{fig:EQ3} and the specific values of the slopes
are included in the caption of the figure.
The growth was, on average, faster for the \emph{bulk} and 
\emph{tangential} seed placements. For the seed placements where the nucleus is
partially immersed in the CO$_2$-rich phase, \emph{3/4 in water} and 
\emph{1/2 in water}, the growth rate was a little slower, which 
indicates that the proximity of the interface decelerates the growth 
of the nucleus. The slowing down of the growth is even more pronounced
for the systems where the seed was partially cut, although in these cases
the comparison is not totally fair because the initial seed size is
smaller.  Since hydrates need water molecules to grow, it is expected that the growth
is faster in the aqueous phase.

\subsection{Simulations of the unconstrained hydrate seeds}

Under the same conditions as we used for the runs presented in the previous section -- 400 bar and 255 K -- we carried out simulations in which the positions of the seeds inserted into the system were not constrained. This time seeds of different sizes were used, with a radius ranging from 1.3 to 2.0 nm. The starting configurations for these runs were equilibrated as described in Supporting Information, where we show that the specific protocol used for the equilibration of the seed interface does not affect
the results. 

\begin{figure*}
\includegraphics[width=0.95\textwidth]{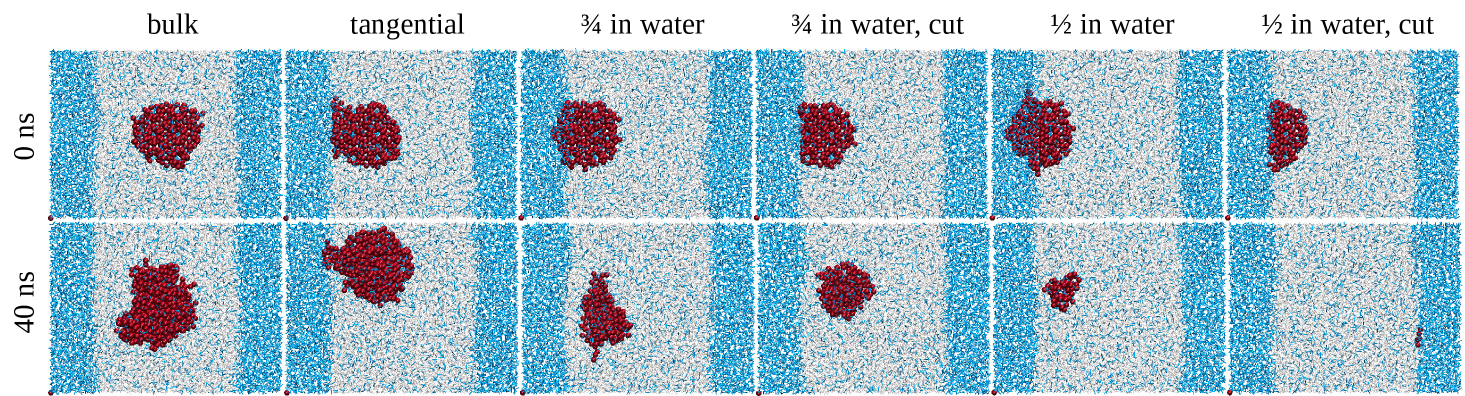}
\caption{Snapshots of the system in unrestrained production runs at 0 and 40 ns. The starting radius of the seeds was 1.7 nm in all cases. The molecules of water that were found to belong to the biggest hydrate cluster (identified as in Ref. \onlinecite{zeron2025homogeneous}) are presented as red spheres. Water molecules in the liquid phase and CO$_2$ molecules are shown as grey and cyan lines, respectively. The graphic was created using VMD\cite{vmd} software.}
\label{fig:md_EQ1}
\end{figure*}

An example of the evolution of the seeds after 40 ns is shown in Figure \ref{fig:md_EQ1}. As can be seen, the seeds originally inserted into the system close to, or immersed in, the CO$_2$-rich phase tend to drift away from the interface into the aqueous solution. 
This observation clashes with the 
hypothesis that nucleation takes place at the interface.

\begin{figure*}
\includegraphics[width=0.90\textwidth]{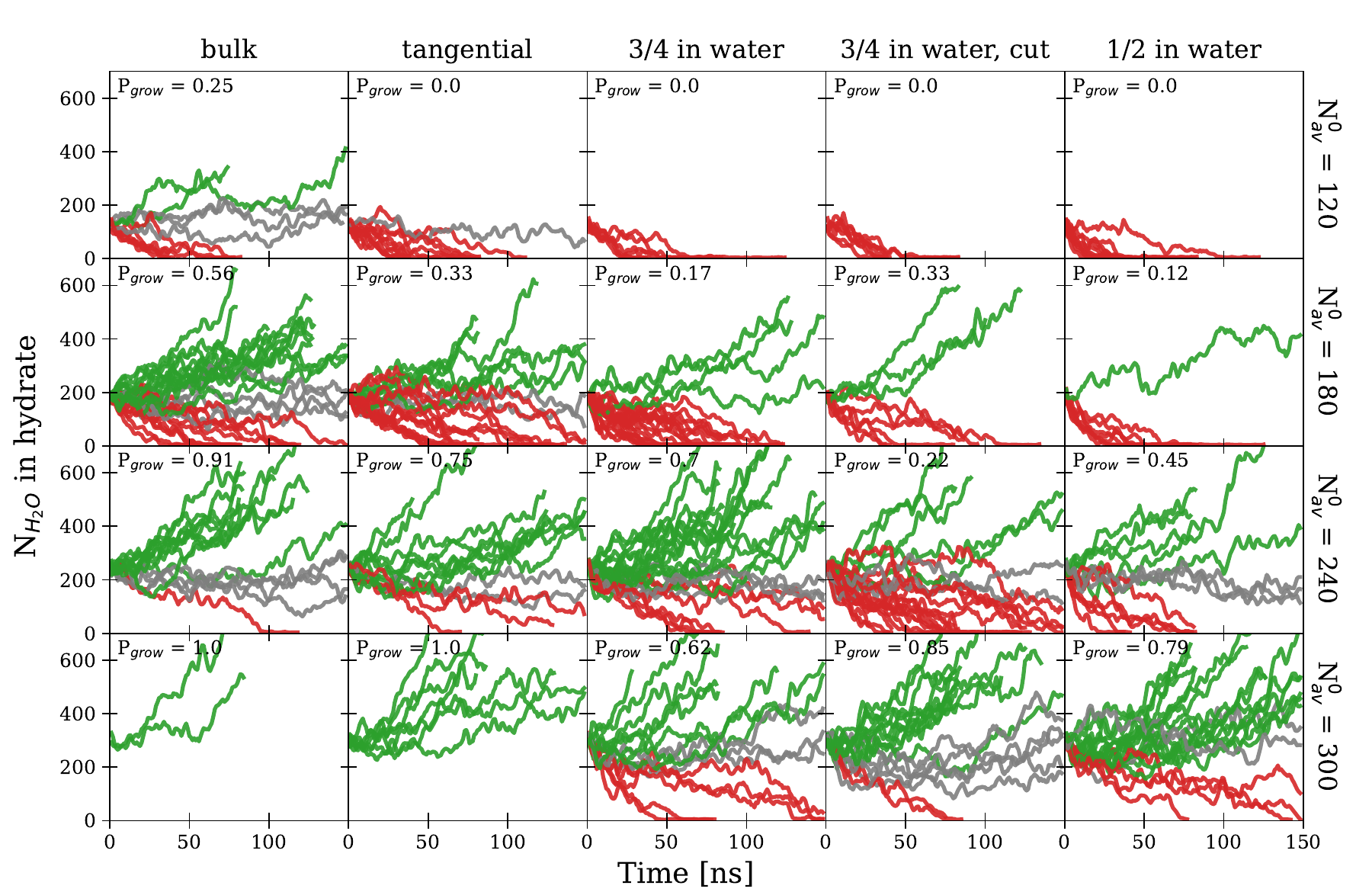}
\caption{Changes in time of the size of the hydrate seed during production runs. Results are shown in separate rows according to the average size of the seed in the first 3 nanosecond of a given production run. The average sizes of the seeds at the start of the runs presented within each row are indicated on the right-hand side of the figure. In the figure, the data for only a few starting seed sizes is shown
(an extended figure with all initial sizes considered is presented
in Supporting Information). The different seed locations/shapes are organized in different columns. Lines in the figure are coloured in red if the seed melts and in green if it grows, while grey lines are used for the runs in which the size of the seed is stable (not changing enough to label it as either "growing" or "melting"). The probability of growing was calculated in each plot based on the number of growing and melting seeds (the runs labelled as "stable" were not included in the calculations).}
\label{fig:seeding_all}
\end{figure*}

Depending on the location, shape and size of the seed at the beginning of the trajectory, the outcomes we observed were different. In some cases, the seeds were growing in most of the runs, in others they were almost always melting. To organize all these results, we analyzed changes in time of the seed sizes. 
The results are presented in Figure \ref{fig:seeding_all}.

In the figure, the different seed locations are organised in columns, 
while rows correspond to a similar size of the seeds at the beginning of the
trajectory (within a range of 20 water molecules around the average value shown on the right side of the figure). For clarity, only a few starting sizes of the seeds are included in Figure \ref{fig:seeding_all} (see Figure S2 in Supporting information for the figure including all runs). The 
\emph{1/2 in water, cut} type was not included in the figure, since in the majority of the runs, regardless the starting seed size, the seeds were melting.
Trajectories are coloured in green and red if the seed size either increases or decreases significantly during the run (the change of the size must be greater than 50\% compared to the starting size of the seed). Trajectories are coloured in grey if the change of the seed size was smaller than 50\%.
The growth probability indicated inside each plot corresponds to the
ratio between the number of green trajectories and the sum of green
and red ones (grey ones are not considered
for calculating the probability).

The growth probability increases as one moves down a given column.
This trend is expected, since the size of the seed from which the trajectories are initiated increases down the column (the larger the seed is, the more 
likely it is that it grows).

By examining a particular row in Fig.~\ref{fig:seeding_all},
one can clearly observe that the probability of growth decreases from left to right,
as the seed is placed closer to the interface.
The finding that the interface hinders the growth of hydrate seeds is one of the main results of this work.

\begin{figure}
\includegraphics[width=0.9\columnwidth]{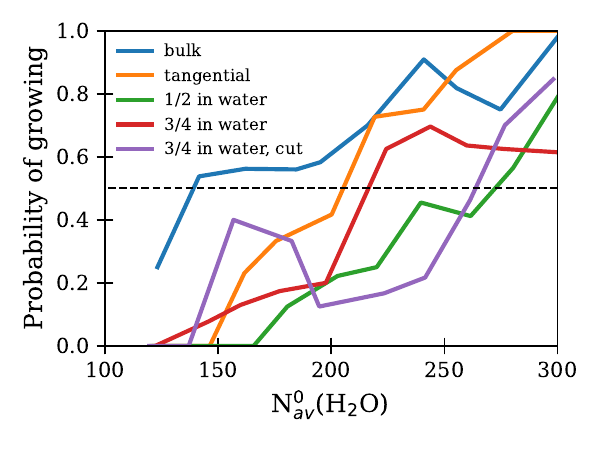}
\caption{Probability of growing of hydrate seeds as a function of the
initial seed size (measured as the number of water molecules belonging to the seed)
for different system types (see legend).}
\label{fig:seeding_probs}
\end{figure}

As mentioned before, within each panel in Fig.~\ref{fig:seeding_all}, we report the corresponding probability of growth.
These probabilities are then plotted in Fig.~\ref{fig:seeding_probs} as a function of the initial seed size for all seed locations and shapes considered.
Although the curves are somewhat noisy due to limited statistics, one can clearly see that the cyan curve - corresponding to \emph{bulk} spherical clusters - stands above the others, indicating that spherical clusters embedded in the bulk molecular environment have the highest tendency to grow.
In contrast, the curves corresponding to the \emph{1/2 in water} and \emph{3/4 in water, cut} configurations (green and purple, respectively) fall below the rest, revealing that these locations are highly unfavourable.
The \emph{1/2 in water, cut} cluster type is even less favourable, but it does not appear in the figure because the growth probability was close to zero for all studied seed sizes.
Overall, the results clearly demonstrate that clusters have a reduced chance of growing when placed near or at the interface.

Furthermore, truncating a cluster so that it presents a planar facet at the
reservoir-solution interface does not promote growth.
This becomes most evident when comparing the \emph{1/2 in water} and \emph{1/2 in water, cut} shapes: whereas clusters in the latter configuration (a hemisphere) never grew, spherical seeds reached a 50\% growth probability for sizes larger than about 250 water molecules (see green curve in Fig.~\ref{fig:seeding_probs}).
The observation that planar faces are disfavoured is consistent with Fig.~\ref{fig:EQ3_picture}, where we show that clusters spontaneously adopt a rounded shape as they grow.
We emphasize that the comparison in Fig.~\ref{fig:seeding_probs} is made at fixed numbers of water molecules in the seed.
Had the comparison instead been carried out at fixed radius, truncated clusters would have had an even lower probability of growth, since they contain fewer molecules than their spherical counterparts (for a fixed radius).

\subsection{Estimation of the nucleation rate for different seed types}

Based on Figure \ref{fig:seeding_probs} it is possible to estimate the critical size of the seed types considered in this work.
The horizontal dashed line in Fig. \ref{fig:seeding_probs} corresponds to a 50\% 
probability. 
We identify the critical size, $N_c$, as that for which 
the growth probability is 50\% (the crossing between the horizontal dashed line and 
a given coloured line). 
The smaller the critical cluster is, the higher the nucleation probability of a certain seed type. 
The estimated critical cluster sizes for most of the seed types examined in this study are reported in Table~\ref{tab:seeding} (for the \emph{1/2 in water, cut} configuration almost all clusters melted and no critical size could be determined, so we presume it is larger than the rest).
From smallest to largest $N_c$ the investigated seed types can be sorted as follows:
\emph{bulk};  \emph{tangential};  \emph{3/4 in water};  \emph{3/4 in water, cut}; \emph{1/2 in water};  \emph{1/2 in water, cut}.
This sequence confirms that bulk (homogeneous) nucleation is more favoured than nucleation 
near or at the interface. Furthermore, it 
indicates that, at the conditions under study, spherical clusters are preferred 
to truncated ones (with the flat face lying on the reservoir-solution interface.) 

Knowing $\Delta \mu_N$, the chemical potential
 difference between the crystal and the solution 
(which we calculated in our previous work for the same conditions, 255 K and 400 bar \cite{Algaba2023a} and is equal to -2.26 k$_B$T), we can then calculate the free energy barrier for nucleation, $\Delta G_c$, using the Classical Nucleation Theory \cite{becker1935kinetische,volmer1926keimbildung,gibbsCNT1,gibbsCNT2} result: $\Delta G_c = |\Delta \mu_N| N_c/2$. 
It is then possible to go further and estimate the nucleation rate using:
\begin{equation}
J = \rho_{L}^{\text{CO}_{2}} \,   Z \, f^{+}_{\text{CO}_{2}} \, \text{exp} \left(\frac{-N_{c}^{\text{CO}_{2}} \, |\Delta \mu_{\text{N}} | }{2 \, k_{B} \, T } \right) 
    \label{cnt}
\end{equation}

\noindent
where $\rho^{\text{CO}_{2}}_{L}$ is the number density of CO$_{2}$ in the liquid phase, $N_c^{\text{CO}_{2}}$ is the number of molecules of CO$_{2}$ in the critical cluster, $Z$ is the Zeldovich factor and $f^{+}_{\text{CO}_2}$ is the attachment rate. 

In order to estimate nucleation rates in our systems, we used the same values of $Z$ and $f^{+}_{\text{CO}_2}$ as in our previous work. \cite{zeron2025homogeneous} $\rho_{L}^{\text{CO}_{2}}$ is simply the number density of CO$_2$ in the aqueous phase.
The parameters used for the calculation of the nucleation rate, as well as the
nucleation rate itself, are reported in Table \ref{tab:seeding}.
Note that the value we obtain for nucleation of \emph{bulk} seeds ($J=10^{23}$ 
m$^{-3}$s$^{-1}$) is consistent within two orders of magnitude 
with the value we published in Ref. \onlinecite{zeron2025homogeneous} of $J=10^{25}$ 
m$^{-3}$s$^{-1}$
(4-5 orders of magnitude is a typical error bar of Seeding nucleation
rate calculations \cite{sanz2013a,niu2019temperature}). 

These estimates, although rough, illustrate that the nucleation rate
decreases by many orders of magnitude from a nucleus emerged in the bulk molecular environment
to other less favourable locations close to or immersed in 
the CO$_2$-rich phase. Even the nucleation
in the second-best seed type, \emph{tangential}, is about 6 orders of magnitude slower than bulk nucleation. When a spherical nucleus is placed with its equatorial line at the interface
(\emph{1/2 in water} location)
one obtains nucleation rates more than 10 orders of magnitude slower than 
in the bulk. We do not find, therefore, any evidence of a preferred nucleation
at the interface in our seeding simulations. 

Let us revisit, however, the spontaneous nucleation simulations 
we performed in our previous work \cite{zeron2025homogeneous} at lower temperatures (245 K
and 250 K) to check whether this finding is affected by the artificial
Seeding of the hydrates cluster in the system. 

\begin{table*}
    \centering
    \caption{Nucleation rate of CO$_{2}$ hydrate in water, $J$, obtained from the seeding simulations at $255\,\text{K}$, $400\,\text{bar}$, for different system types, and parameters used in order to obtain nucleation rates. The size of the critical cluster was estimated as number of water molecules belonging to the hydrate, with the use of linear combination of $\overline{q}_3$ and $\overline{q}_{12}$ order parameters.\cite{zeron2025homogeneous} The number of CO$_2$ molecules in the critical cluster was then calculated by dividing $N_{c}^{\text{H}_{2}\text{O}}$ by a factor of 5.75, which corresponds to the water to CO$_2$ ratio in a crystalline sI hydrate. Note that $N_{c}^{\text{CO}_{2}}$ is equivalent to the number of cages given that we cut the seeds from a thermalised, fully occupied CO$_2$ hydrate lattice. The estimated uncertainty for the critical nucleus size is $\pm$ 20 water molecules (estimated from the bin width used to classify initial seed sizes), leading to relative uncertainties of about 5–10\% in the nucleation barrier and to  nucleation rate uncertainties of approximately 4–5 orders of magnitude.}
    \vspace{0.5cm}
    \begin{tabular}{lcccccccccc}
    \hline
    \hline
          & \,\, & bulk & \,\, &  tangential & \,\, & 3/4 in water & \,\, & 3/4 in water, cut & \,\, & 1/2 in water \\
    \hline 
        \\[-0.3cm]
        $N_{c}^{\text{H}_{2}\text{O}}$ & \,\, &  140 & \,\, & 205 & \,\, & 217 & \,\, & 264 & \,\, & 272 \\
        $N_{c}^{\text{CO}_{2}}$ & \,\, & 24.3 & \,\, & 35.7 & \,\, & 37.7 & \,\, & 45.9 & \,\, & 47.3  \\
        $\Delta G_{c}/(k_BT)$ & \,\, & 27.5  & \,\, & 40.3 & \,\, & 42.6 & \,\, & 51.9 & \,\, & 53.5 \\
        $J\,(\text{m}^{-3}\,\text{s}^{-1})$ & \,\, & 1 $\times 10^{23}$ & \,\, & 4 $\times 10^{17}$ & \,\, & 4 $\times 10^{16}$ & \,\, &  4 $\times 10^{12}$ & \,\, & 8 $\times 10^{11}$ \\
    \hline
        \\[-0.32cm]
        $Z$ & \,\,  & \multicolumn{9}{c}{0.08}  \\
        $f^{+}_{\text{CO}_{2}}\,(\text{s}^{-1})$ & \,\, & \multicolumn{9}{c}{6.54 $\times 10^{8}$} \\
        $\rho_{L}^{\text{CO}_{2}}\,(\text{m}^{-3})$ & \,\,  & \multicolumn{9}{c}{2.6 $\times 10^{27}$}  \\[0.5mm]
        
    \hline
    \hline        
    \end{tabular}
    \label{tab:seeding}
\end{table*}

\label{J_calc}

\subsection{Unseeded spontaneous nucleation}

In previous work, \cite{zeron2025homogeneous} we demonstrated that at 245 K and 250 K under the pressure of 400 bar, spontaneous hydrate nucleation occurs within hundreds of nanoseconds.
Furthermore, we found that the nucleation time is identical in a CO$_2$-saturated bulk aqueous solution and in a two-phase system, where a slab of the aqueous solution
is saturated with CO$_2$ by contact with a reservoir through a flat interface (note that the CO$_2$ concentration in the aqueous phase is
the same in one-phase and two phase-systems).
This finding clearly shows that, at least at these two temperatures, the presence of the interface does not accelerate nucleation.
Moreover, this result also remarks that 
bulk-like behaviour can be found in the aqueous slab of 
the two-phase system, as illustrated by the density profiles 
shown in Fig. \ref{fig:densprof}.

\begin{figure}
    \centering
    \includegraphics[width=0.9\columnwidth]{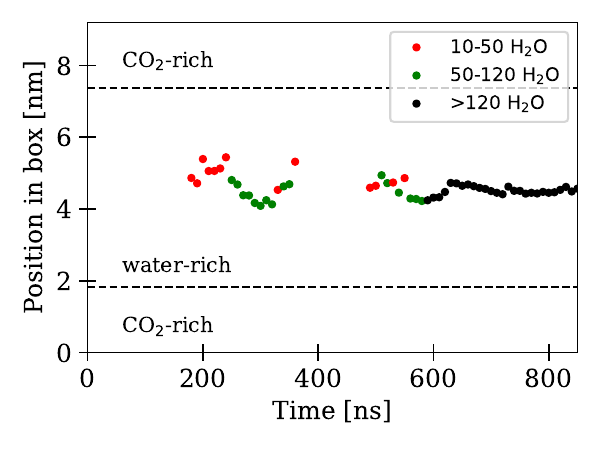}
    \caption{Dots indicate the coordinate perpendicular to the interface of the centre of mass of the 
    of the largest hydrate nucleus in a spontaneous nucleation 
    trajectory at 250 K and 400 bar.  
    The average location of the interfaces is indicated with the dashed horizontal lines. Different dot colours represent different size ranges of the nucleus 
    (see legend), in terms of number of water molecules. Only sizes larger than 10 water molecules were considered. }
    \label{fig:locationspontaneous}
\end{figure}

To corroborate this result, we analyse in this work the location of hydrate nuclei
that appear spontaneously at 250 K in the two-phase system. 
According to the Seeding analysis performed in this work and to the 
spontaneous nucleation rate calculations conducted in Ref. \onlinecite{zeron2025homogeneous}, it is
expected that these nuclei do not appear at the interface. 
In Fig. \ref{fig:locationspontaneous} we show the centre of mass location of the nucleus
along time using a certain colour code depending on the nucleus size (see legend)
and for a selected spontaneous nucleation trajectory at 250 K. 
The horizontal dashed lines indicate the average location of the interface between the CO$_2$-rich
and the water-rich phases. 
Clearly, the nucleus does not appear close to any of the two interfaces, 
corroborating our main result that nucleation does not take place at
the interface. 
A similar behaviour was observed in all nucleating trajectories.
In Ref. \onlinecite{hu2022three}, where spontaneous CH$_4$ hydrate nucleation was
studied in systems containing a CH$_4$ bubble embedded in 
the aqueous solution, they observed the appearance
of the hydrate nucleus away from the bubble-aqueous solution interface, 
in agreement with what we see here for planar interfaces. 

\begin{figure*}
  \centering
  \includegraphics[width=0.9\linewidth]{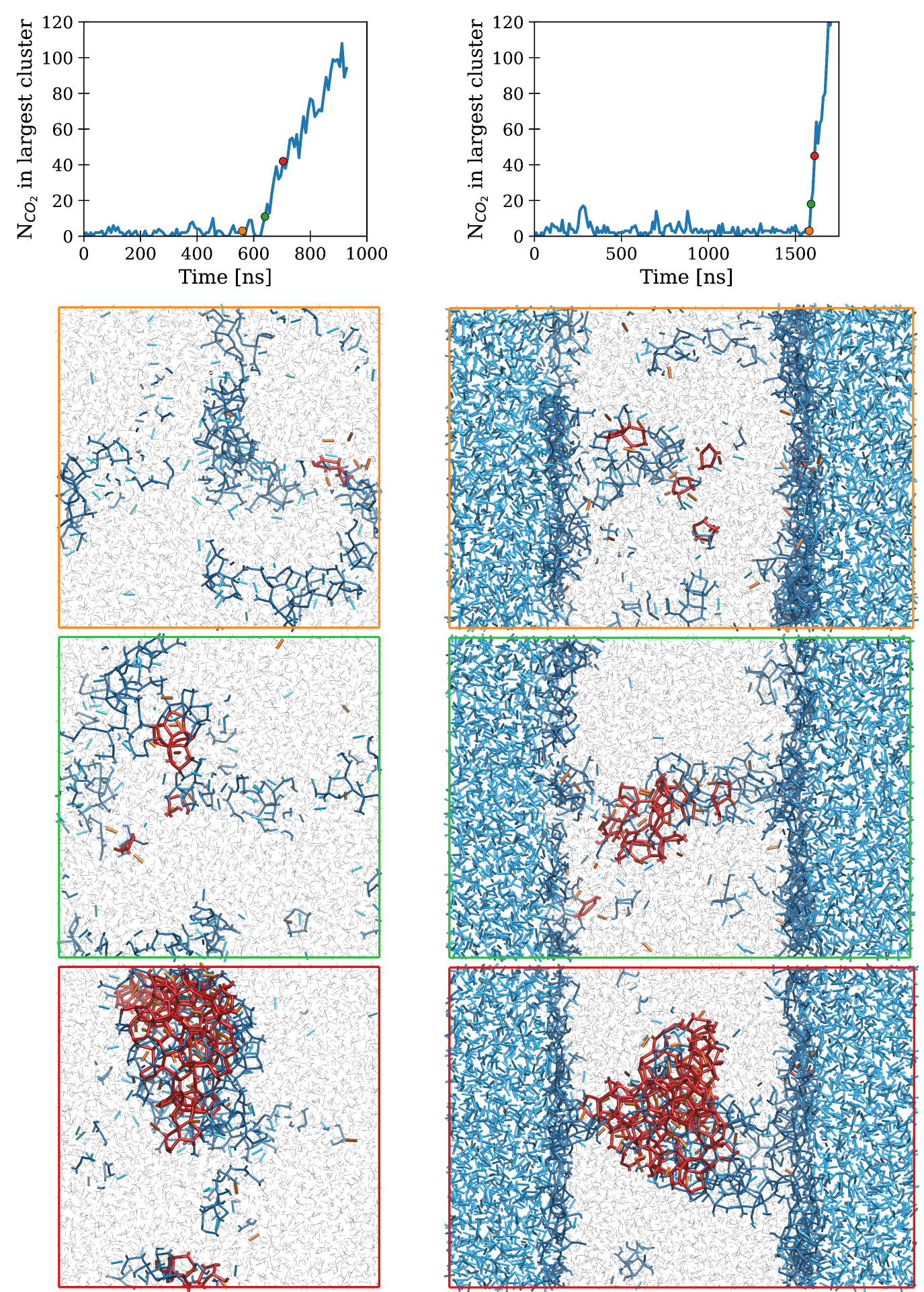}
 \caption{Top panel: size of the largest hydrate cluster in time in representative spontaneous nucleation trajectories in the one-phase (left column) and the two-phase systems (right column) at 400 bar and 250 K. For three moments in each trajectory, indicated by coloured dots in the top plots, snapshots framed in the corresponding color are shown below. The clusters of hydrate recognized by the MCG-3 order parameter are marked with orange (CO$_2$ molecules) and red (water molecules) bonds. Areas of locally higher density of CO$_2$ are also shown, as light blue (CO$_2$ molecules) and dark blue (water molecules) bonds. The areas of high CO$_2$ density were found by looking for molecules of CO$_2$ that have 11 or more neighbouring molecules of CO$_2$ (within distance of 0.9 nm).The graphic was created using VMD\cite{vmd} software.}
\label{fig:co2_local_density}
\end{figure*}

In Fig. \ref{fig:co2_local_density} we show an analysis of the spontaneous nucleation path in bulk and two-phase
systems. In the top part of the figure we show the number of CO$_2$ molecules
in the largest hydrate cluster identified with the MCG-3 algorithm.\cite{mcg3} Below these plots, three snapshots of the systems at times specified by colored dots are presented. 

For this particular analysis we chose the MCG-3 order parameter because it allows to
identify the molecules of CO$_2$ belonging to the hydrate seed, unlike the 
linear combination of $\overline{q}_3$ and $\overline{q}_{12}$ order parameters \cite{zeron2025homogeneous} used in Table \ref{tab:seeding}  which refers to water molecules. The MCG-3 order parameter labels a molecule of CO$_2$ as hydrate if a set of geometrical criteria is met. Firstly, a molecule of CO$_2$ has to have a neighboring molecule of CO$_2$ within 9 \AA~ and this pair of CO$_2$ molecules has to have at least 5 molecules of water "shared" between them. "Shared" water molecules, as described in Ref. \onlinecite{mcg3}, correspond to molecules of water inside an area limited by two cones located along the line connecting the two CO$_2$ molecules and starting in the carbon atoms of CO$_2$ molecules (semi-vertical angles of the cones are equal to 45 degrees). Secondly, a molecule of CO$_2$ can be labeled as hydrate only if it forms at least three pairs described in the previous step.  

In the snapshots of Figure \ref{fig:co2_local_density}, molecules of CO$_2$ labeled as hydrate by the MCG-3 order parameter and the molecules of water "shared" by pairs of CO$_2$ molecules in the hydrate are highlighted in orange and red, respectively. Additionally, we searched for regions in the system where the concentration of CO$_2$ was locally higher than the average. For that purpose, we found molecules of CO$_2$ that have at least 11 neighboring CO$_2$ molecules within 9 \AA~ (the same distance criterion as for the MCG-3 order parameter) and highlighted them in cyan. For each pair of CO$_2$ molecules found this way, we also highlighted (in dark blue) the molecules of water "shared" between them. 

It is worth noting that MCG-3 does not label molecules of water as either hydrate or liquid. 
However, it enables identification of regions with a high local 
CO$_2$ concentration and CO$_2$ molecules in a hydrate-like arrangement.
Thus, it is a suitable order parameter for our purpose of studying 
from a qualitative perspective 
the emergence of hydrates in regions of locally high CO$_2$ concentration.  
For obtaining quantitative estimates of the nucleation rate (Section \ref{J_calc}) we chose the linear combination of $\overline{q}_3$ and $\overline{q}_{12}$ proposed in Ref. \onlinecite{zeron2025homogeneous}, which was previously shown to provide seeding estimates of $J$ in good agreement with spontaneous nucleation \cite{zeron2025homogeneous}.

As can be seen in the top part of Figure \ref{fig:co2_local_density}, in both system types - bulk and containing planar interface with a CO$_2$ reservoir - there is an induction period of a few hundreds of nanoseconds where small
clusters of hydrate form and redissolve until a stochastic fluctuation gives rise to a 
critical cluster 
that quickly grows.

In the snapshots we can see a similar nucleation mechanism in both cases.
The presence of cyan-blue clustered regions before nucleation indicates that there is strong CO$_2$ density heterogeneities. Obviously, in the two-phase system the interface
corresponds to one of these regions, but there are also clear CO$_2$ density fluctuations
in the bulk aqueous phase. Interestingly, hydrate clusters always appear within 
a bulk high CO$_2$ density regions. 
Thus, we envisage a mechanism where CO$_2$ density fluctuations facilitate the
emergence of hydrate clusters (in fact, spontaneous CH$_4$ nucleation 
occurs when the solution is supersaturated with methane \cite{Wu2009}). 
Similar observation of nucleation in regions of high local guest molecule
concentration has already been reported in previous works. \cite{vatamanu2010observation,Jacobson2010a,li2020unraveling}
However, our analysis reveals that 
 these fluctuations are present both close to the CO$_2$-aqueous solution interface and in the bulk 
aqueous phase, but the proximity of the interface disrupts the nucleation of the hydrate, 
making the bulk aqueous solution the preferred location for nucleation.

\section{Discussion}
Our simulations clearly show that nucleation preferentially occurs in 
the bulk. Does our work imply that hydrate nucleation also takes place in the bulk in real systems? Experiments observe hydrate appearance and growth at the interface. \cite{adamova2018visual,li2024dependence,jeong2022gas,maeda2015nucleation,sloan2007clathrate} 
Here we provide some speculative
hypotheses to reconcile our simulation results
with experimental observations.

One possibility is 
that nucleation takes place in the bulk (homogeneous nucleation) and then the nucleus attaches to 
the interface, where it subsequently grows. 
However, this scenario seems unlikely because experiments generally observe hydrate
nucleation at temperatures a few kelvin below the dissociation temperature, \cite{Maeda2018,Maeda2019,li2024dependence,adamova2018visual,metaxas2019gas,jeong2022gas}
and several simulation studies estimate that homogeneous nucleation  
is not possible at such a small supercooling. \cite{Knott2012a,zeron2025homogeneous} 

Where, then, is nucleation actually taking place?
One possibility is that 
there is a crossover from bulk (homogeneous) to interfacial (heterogeneous) nucleation as temperature increases.
In our simulations, bulk nucleation of CO$_2$ hydrates is faster at 245, 250 and 255 K. 
However, the homogeneous
nucleation rate drops rapidly with increasing temperature. In contrast, the 
interfacial nucleation rate may have a weaker temperature dependence. 
This opens the possibility that at higher temperatures interfacial nucleation 
becomes dominant. 

Another plausible scenario is that nucleation happens in the contact line 
between the aqueous solution, the hydrate-former rich phase and the cell walls. 
This possibility has been proposed in several papers \cite{maeda2015nucleation,jeong2022gas,Maeda2019,stoporev2018visual}
and may also explain
why 
higher supercoolings are typically observed when hydrate formation occurs in droplets suspended without contact with solid walls. \cite{maeda2015nucleation,jeong2022gas,davies2009studies}
Additionally, the presence of impurities, both in the bulk or trapped at the interface, may also increase the
temperature at which hydrate formation is observed. 
Simulations show that solid surfaces can indeed promote hydrate nucleation.
\cite{zhang2025temperature,bai2015properties}

We emphasize that our work focuses on the early nucleation stage. Rapid growth may require a fast supply of hydrate-former molecules - CO$_2$ in our case - from the hydrate-former-rich phase. Consequently, nucleation and early growth may occur in the bulk molecular environment (homogeneous nucleation), whereas sustained growth may require the post-critical nucleus to be in close proximity to the interface. The time and length scales required to investigate such growth processes are beyond those accessible to our molecular simulations.

Our work appears to clash with a recent simulation study where
nucleation is reported to take place at the interface at high supercooling. \cite{zhang2025temperature}
A possible difference between our work and that of Ref. \onlinecite{zhang2025temperature}
is that they use H$_2$S instead of CO$_2$ as hydrate former.
For the model used in this work for CO$_2$ the solubility of CO$_2$ decreases as the temperature increases \cite{zeron2025homogeneous}, whereas in Ref. \onlinecite{zhang2025temperature} the solubility of H$_2$S increases
as \emph{T} increases (see Fig. 4d in Ref. \onlinecite{zhang2025temperature}). Experimentally the solubility of both guest molecules decreases
as T increases, at least under low pressures.\cite{sander2015compilation} 
Further studies are needed
to clarify whether there are fundamental differences between H$_2$S and CO$_2$
hydrate nucleation. 

In any case, our results challenge the commonly held view that hydrate nucleation universally occurs at interfaces. At high supercooling, bulk (homogeneous) nucleation may in fact dominate. More work is needed to understand the nucleation mechanism at moderate supercooling, where most experimental observations are made.

A promising direction is the direct comparison of nucleation rates from simulations and experiments. This quantity provides a natural bridge between both approaches, allowing validation of simulations, which offer access to molecular-level details that are often inaccessible experimentally. For ice nucleation, the nucleation rate has already 
been successfully compared  between simulations (using the same water model as in the present work) and experiments \cite{espinosa2018homogeneous,niu2019temperature} 
(even though there is still an open debate over the competition between 
bulk and surface ice nucleation in small water drops \cite{sun2024surface,haji2017computational}).
Unfortunately, for hydrates, very few experimental measurements of nucleation rates are available. In
contrast, there are already several simulation studies 
where hydrate nucleation rates have been calculated. Most of them focus 
on the bulk, but often using aqueous solutions with unrealistically high hydrate-former concentrations. \cite{Walsh2011a,Warrier2016a,zeron2025homogeneous,Grabowska2022b,Knott2012a}

 Our simulations suggest that at sufficiently high supercooling, homogeneous (bulk) nucleation should outpace interfacial nucleation. This regime may offer a valuable opportunity to directly measure homogeneous nucleation rates and compare them with existing simulation predictions, thereby laying a stronger foundation for the molecular-level understanding of hydrate nucleation.

\section{Summary and conclusions}

We employ molecular dynamics simulations of a realistic water–CO$_{2}$ model to examine the formation of CO${_2}$ hydrates in supercooled, CO$_{2}$-saturated aqueous solutions at 400 bar. Most simulations are carried out at 255 K, corresponding to a supercooling of 35 K below the hydrate–solution–CO${_2}$ triple point.\cite{Algaba2023a} Hydrate seeds are introduced at different positions relative to the interface to monitor their growth. Interestingly, we find that hydrate nucleation proceeds more rapidly in the bulk than near the interface, challenging the conventional view. We also explore spontaneous nucleation pathways (without seeding) at 250 K, a higher supercooling that enables homogeneous hydrate nucleation within our simulation timescale. In this regime, nucleation occurs preferentially in regions with locally enhanced CO$_{2}$ concentration due to thermal fluctuations; these regions arise in the bulk and are not associated with the interface. Our findings highlight the need for careful investigation of hydrate nucleation mechanisms, a process of critical relevance to flow assurance in the oil industry and to natural environmental systems. Further work is required to reconcile these observations with experimental evidence on hydrate nucleation and growth.

\section{Supporting Information}
Additional details on the equilibration protocol used in this study, along with an extended version of Figure \ref{fig:seeding_all} showing hydrate size evolution across all simulations (PDF). 
\section{Acknowledgements}
J. G. gratefully acknowledge Polish high-performance computing infrastructure PLGrid (HPC Center: ACK Cyfronet AGH) for providing computer facilities and support within computational grant no. PLG/2025/018076. Computations were carried out using the computers of Centre of Informatics Tricity Academic Supercomputer \& Network. C.~Vega, E.~Sanz, and S.~Blazquez acknowledge the funding from project PID2022-136919NB-C31 of Ministerio de Ciencia, Innovación y Universidades. 

\section{References}

\end{document}